\documentclass{article}
\usepackage{spconf,amsmath,graphicx}

\usepackage{url}            
\usepackage{booktabs}       
\usepackage{amssymb}
\usepackage[normalem]{ulem}
\useunder{\uline}{\ul}{}
\usepackage{bibspacing}

\title{ISPA: Inter-Species Phonetic Alphabet for Transcribing Animal Sounds}
\name{Masato Hagiwara \quad Marius Miron \quad Jen-Yu Liu}
\address{Earth Species Project}
\begin{document}
%
\maketitle
\begin{abstract}
Traditionally, bioacoustics has relied on spectrograms and continuous, per-frame audio representations for the analysis of animal sounds, also serving as input to machine learning models. Meanwhile, the International Phonetic Alphabet (IPA) system has provided an interpretable, language-independent method for transcribing human speech sounds. In this paper, we introduce ISPA (Inter-Species Phonetic Alphabet), a precise, concise, and interpretable system designed for transcribing animal sounds into text. We compare acoustics-based and feature-based methods for transcribing and classifying animal sounds, demonstrating their comparable performance with baseline methods utilizing continuous, dense audio representations. By representing animal sounds with text, we effectively treat them as a ``foreign language,'' and we show that established human language ML paradigms and models, such as language models, can be successfully applied to improve performance.
\end{abstract}

\section{Introduction}

Traditionally, bioacoustics has relied on spectrograms and sonograms---visual representations depicting a sound's frequency and amplitude over time---for the visualization and analysis of animal sounds. Machine learning approaches in bioacoustics typically utilize continuous, per-frame audio representations as input~\cite{stowell2022computational}. In contrast, human speech sounds are often transcribed into textual representations, commonly a sequence of phones and phonemes using the International Phonetic Alphabet (IPA). The transcription of human speech in a standardized, concise, and interpretable manner opens avenues for diverse research and downstream tasks, such as linguistic analysis and speech understanding. In this paper we aim at answering the following research question: can we devise a similarly standardized, interpretable yet concise transcription system for animal sounds?

Animal sounds are often transcribed through {\it onomatopoeia} ---words that imitate the sound---in human speech. Many bird resources adopt this format, such as using 'caw caw' for crows and 'chick-a-dee-dee-dee' for chickadees. However, onomatopoeic representations are heavily influenced by language and culture~\cite{bredin1996onomatopoeia}, rendering them unsuitable as a standardized text representation.

There have been only a few attempts to transcribe animal vocalizations. Huang et al. proposed ShibaScript~\cite{huang2023transcribing}, a system for phonetically transcribing domestic Shiba Inu barks. They designed a set of IPA symbols for transcription through clustering and manual inspection. However, it's important to note that their system is highly specific to one species of dogs.

In this paper, we introduce ISPA (Inter-Species Phonetic Alphabet)\footnote{\url{https://github.com/earthspecies/ispa}} for transcribing animal sounds into text. Our objective is to create a transcription system that is 1) precise (retaining the information from the original audio), 2) concise (expressing the information with minimal tokens), and 3) interpretable (understandable by humans). While traditional per-frame continuous representations of audio offer precision, they lack conciseness and interpretability. Onomatopoeia, on the other hand, are concise but may lack precision or interpretability outside the speakers of a particular language. We aim to strike a balance among these three criteria.

Transcribing animal sounds into text offers several advantages. Beyond obvious benefits such as conciseness and interpretability, we can leverage recent advancements in the language field, including large language models (LLMs)~\cite{zhao2023survey}. This facilitates tasks like pretraining, fine-tuning, generation, and multi-modal training, treating ISPA-transcribed sounds as a form of ``foreign language'' for natural language models. Furthermore, reproducing the original audio should also be a feasible task, akin to text-to-speech (TTS)~\cite{ren2022fastspeech2}.

We acknowledge the existence of machine learning methods like ES-KMeans~\cite{Kamper2017eskmeans} and SlowAEs~\cite{dieleman2021slowae}, which convert audio into variable-rate, discrete representations. However, these methods are not straightforward to train and have not gained widespread adoption.

\section{Method}

\subsection{ISPA-A — Acoustics-based transcription}

\begin{figure*}[t]
\begin{center}
\includegraphics[scale=0.33]{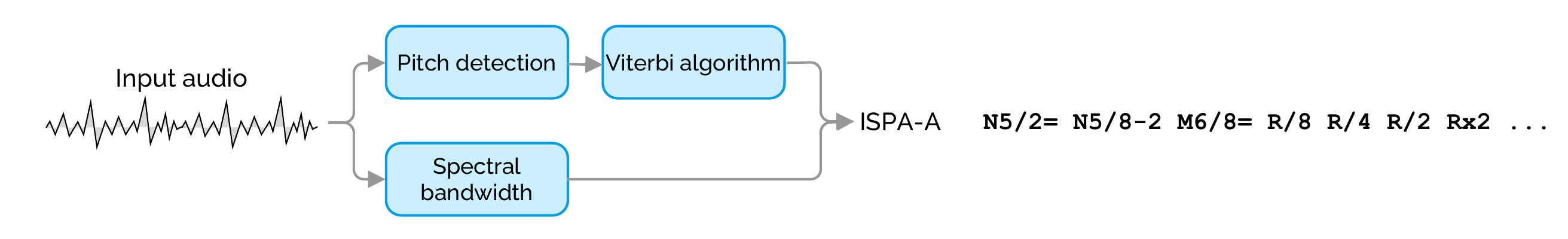} 
\vspace{-1.5em}
\caption{Overview of acoustics-based transcription (ISPA-A)}
\vspace{-1.5em}
\label{fig:ispaa}
\end{center}
\end{figure*}

\begin{figure*}[t]
\begin{center}
\includegraphics[scale=0.33]{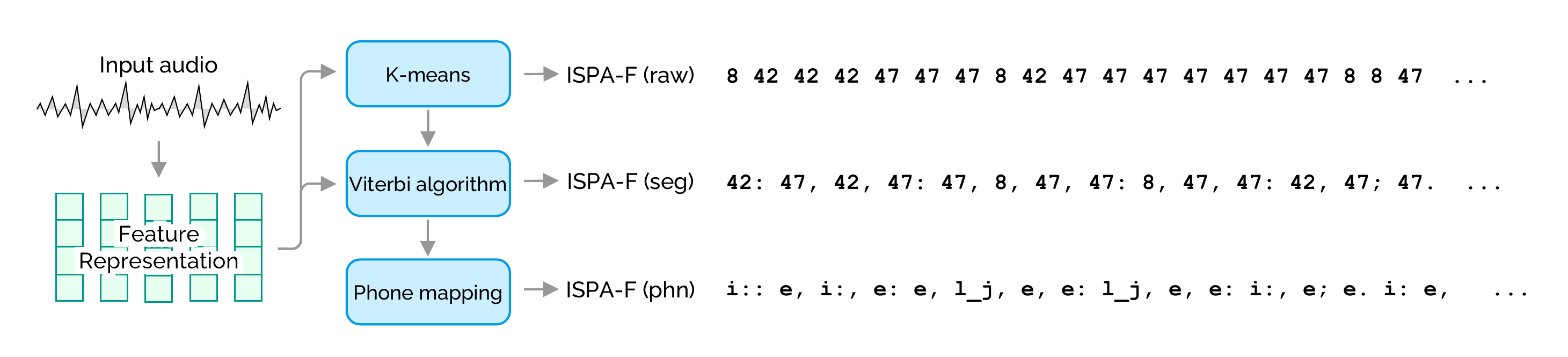} 
\vspace{-1.5em}
\caption{Overview of feature-based transcription (ISPA-F)}
\vspace{-1.5em}
\label{fig:ispaf}
\end{center}
\end{figure*}

The first method we discuss here relies on the acoustic properties of the input audio. This approach is heavily influenced by automatic music transcription (AMT)~\cite{benetos2019amt}, revolving around the identification of a sequence of musical notes that best matches the input audio in terms of pitch and duration. However, there are key distinctions from traditional AMT settings—pitch changes during notes are more frequent in animal sounds, and the transcription should also encode timbre information (the quality of the sound, corresponding to the type of instrument), which may be crucial for animals. While small pitch differences (e.g., a half-note distinction between C and C\#) are crucial in AMT, they might not have the same impact on animal communication for most species.

We designed the acoustics-based ISPA (referred to as ISPA-A) to accurately encode these factors. Specifically, ISPA-A transcription is a sequence of tokens where each token encodes the following acoustic properties of the corresponding audio:
\begin{itemize}
    \setlength\itemsep{-0.5em}
    \item Spectral bandwidth~\cite{klapuri2006signal}: {\tt U} (narrowest), {\tt N}, {\tt M}, {\tt W}, {\tt X} (widest)
    \item Pitch (in MIDI note octave number) or rest ({\tt R} when $<$ -50dB)
    \item Length: {\tt /32} (32nd note), {\tt /16}, {\tt /8}, {\tt /4}, {\tt /2}, $\varnothing$ (whole note), {\tt x2}, {\tt x4}, based on 60 bpm
    \item Pitch slope: {\tt -3} (50\% decrease in pitch), {\tt -2}, {\tt -1}, {\tt =} (no change), {\tt +1}, {\tt +2}, {\tt +3} (100\% increase in pitch)
\end{itemize}

For example, the first ISPA-A token in Figure~\ref{fig:ispaa}, {\tt N5/2=} means that the segment has a narrow spectral bandwidth (resembling a more sine-wave like sound), a pitch in the fifth octave range ({\tt 5}, approx. 500-1,000 Hz), the length of a half note ({\tt /2}, 2 seconds at 60 bpm), and an unchanged pitch ({\tt =}).

Given an audio input, we first apply PESTO~\cite{riou2023pesto} for pitch detection, providing estimated pitch in hertz per specified time frame with confidence. We set each frame to 31.25 ms, aligning with the duration of a 32nd note at 60 bpm. The raw pitch estimation results are challenging to use directly for transcription due to the abundance of data points and its continuous nature. Therefore, we applied a Viterbi algorithm, commonly used in music analysis and pitch estimation~\cite{salamon2014melody}. This algorithm converts the per-frame pitch estimation results into a sequence of variable-length 'segments' that encapsulate both pitch and duration information. Specifically, the algorithm identifies the optimal segment sequence $S = {s_1, ..., s_N}$ by minimizing the cost function $C(S)$ defined below:
\begin{eqnarray}
    C(S) &=& \sum_{i=1}^N \left[ D(s_i) + \lambda LP(s_i) \right] \\
    D(s) &=& \sum_{t=st(s)}^{ed(s)} | f_P(t) - f_s(t) | \\
    LP(s) &=& \frac{1}{1 + \log len(s)}
\end{eqnarray}
Here, the cost function $C(S)$ is a per-token sum of the distance $D()$ and the length penalty $LP()$. The distance is defined as the absolute difference between the estimated pitch provided by PESTO $f_P(t)$ and the estimated pitch within the segment $f_s(t)$ at time $t$ (which may change mid-segment due to the slope), summed from the start of the segment $st(s)$ to its end $ed(s)$. The length penalty term serves to penalize short segments. These two terms maintain a trade-off relationship. Without the penalty term, optimizing the distance would be trivial---simply finding a sequence of one-frame segments that perfectly matches the estimated pitch. In essence, the algorithm aims to identify a sequence that comprises long segments while fitting the pitch estimation results as closely as possible. The Viterbi algorithm, a dynamic programming algorithm, efficiently determines the optimal $S^*$ with linear complexity relative to the input length.

\subsection{ISPA-F — Feature-based transcription}

We also explore a second approach, referred to as ISPA-F (feature-based ISPA), for transcribing animal speech. This method begins with audio features, which are continuous multi-dimensional representations, and transforms them into a sequence of discrete and interpretable segments. By utilizing pretrained models such as AVES~\cite{hagiwara2023aves} (as described later), this method can harness robust representations that have demonstrated efficacy in encoding animal vocalizations.

The core of ISPA-F involves a process where the input sequence of per-frame continuous feature representations is transformed into a sequence of variable-length discrete segments. While there are various approaches to achieving this, we employed k-means clustering for discretization and a Viterbi algorithm for segmentation. Other viable options include ES-KMeans~\cite{Kamper2017eskmeans} and VQ-VAE~\cite{oord2017neural}. Specifically, we first run k-means clustering on a generic audio dataset (using FSD50k~\cite{fonseca2020fsd50k} in our experiments) represented by features to obtain a set of cluster centroids. Subsequently, for a given input audio represented by features, we run a Viterbi algorithm, similar to the one used for ISPA-A, to find the optimal segment sequence $S^*$. The only difference is the choice of the distance function for each segment $D(s)$, defined as:
\begin{eqnarray}
    D(s) &=& \sum_{t=st(s)}^{ed(s)} || f(t) - c(s) ||^2
\end{eqnarray}
where $f(t)$ is the audio feature vector at time $t$ and $c(s)$ is the feature vector that corresponds to the cluster centroid of segment $s$. In ISPA-F, we use punctuation marks similar to English to denote the segment length in frames, i.e., $\varnothing$ = 1, {\tt ,} = 2, {\tt :} = 5, {\tt ;} = 10, {\tt .} = 20, and {\tt ..} = 50.

\begin{figure}[t]
\begin{center}
\includegraphics[scale=0.47]{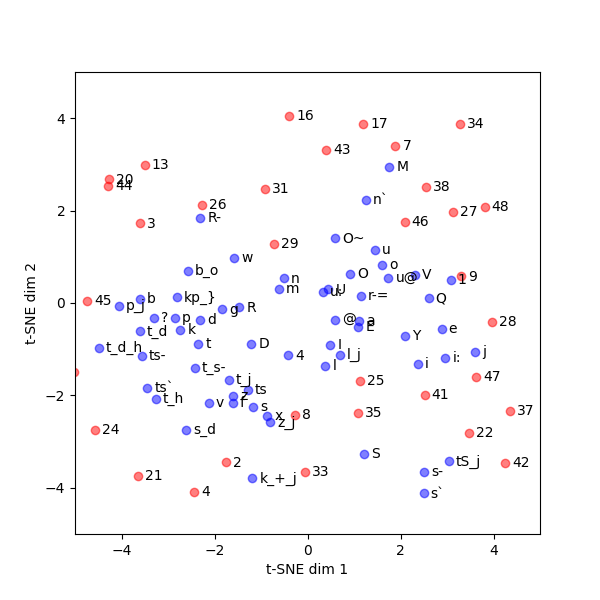}
\vspace{-1.5em}
\caption{t-SNE plot of phone vectors (blue dots) and centroid vectors (red dots)}
\vspace{-1.5em}
\label{fig:phone_cluster}
\end{center}
\end{figure}

ISPA-F can capture acoustic properties that may not be modeled by ISPA-A. However, it has the drawback that the embeddings are not directly interpretable. To achieve interpretability, we implemented an additional step to convert these IDs into human-readable symbols. As a proof-of-concept, we chose to use IPA for describing cluster sounds. First, we use allosaurus~\cite{li2020universal} to transcribe a generic audio dataset (specifically, FSD50K, including various sounds beyond human speech) into the IPA format represented in X-SAMPA~\cite{wells1995xsampa}. Next, we obtained a mean feature vector per phone using timestamp information from allosaurus. Finally, in order to find a one-to-one mapping between phone vectors and centroid vectors, minimizing the distance between them, we solve this as a linear sum assignment problem (via scipy's {\tt scipy.optimize.linear\_sum\_assignment}). Figure~\ref{fig:phone_cluster} illustrates a t-SNE plot of phone vectors and centroid vectors obtained using AVES. While there is some overlap (e.g., consonant sounds like 's,' 'k,' 'z,' and 'M'), centroid vectors are dispersed across a broader area in the latent space. This suggests that the ISPA-F representation encompasses a more diverse range of sounds beyond human speech phones.

Figure~\ref{fig:ispaf} shows the overview of ISPA-F. We compare three sub-variations of ISPA-F—the sequence of per-frame cluster IDs without the Viterbi algorithm ({\tt raw}), the sequence of variable-length segments after the Viterbi algorithm ({\tt seg}), and the sequence of phone-mapped segments ({\tt phn}).

\section{Experiments}

\begin{table*}[!t]
\begin{center}
{\small
\begin{tabular}{@{}lrrrrrrr@{}}
\toprule
                  & \multicolumn{1}{l}{esc50} & \multicolumn{1}{l}{watkins} & \multicolumn{1}{l}{bats} & \multicolumn{1}{l}{cbi} & \multicolumn{1}{l}{hbdb} & \multicolumn{1}{l}{dogs} & \multicolumn{1}{l}{tokens / sec} \\ \midrule
IPA               & 0.058                     & 0.043                       & 0.139                    & 0.003                   & 0.351                    & 0.314                    & 1.2                              \\
ISPA-A            & 0.207                     & 0.394                       & 0.224                    & 0.003                   & 0.695                    & 0.458                    & 2.7                              \\
ISPA-F, MFCC, raw & 0.350                     & 0.653                       & {\bf 0.460}                    & 0.003                   & 0.731                    & 0.588                    & 50.0                             \\
ISPA-F, MFCC, seg & 0.340                     & {\bf 0.657}                       & 0.375                    & 0.003                   & {\bf 0.737}                    & {\bf 0.597}                    & 8.3                              \\
ISPA-F, MFCC, phn & 0.345                     & 0.620                       & 0.381                    & 0.003                   & 0.735                    & 0.583                    & 8.3                              \\
ISPA-F, AVES, raw & {\bf 0.420}                     & 0.569                       & 0.459                    & 0.003                   & 0.328                    & 0.506                    & 49.8                             \\
ISPA-F, AVES, seg & 0.358                     & 0.431                       & 0.302                    & {\bf 0.004}                   & 0.725                    & 0.475                    & 3.8                              \\
ISPA-F, AVES, phn & 0.380                     & 0.454                       & 0.296                    & 0.003                   & 0.715                    & 0.465                    & 3.8                              \\ \bottomrule
\end{tabular}
}
\end{center}
\vspace{-1.0em}
\caption{Main results with the RoBERTa base model. Accuracy is an average of three evaluation runs. The token / sec results are measured on the ESC50 training set.}
\label{table:results-base}
\end{table*}

\begin{table*}[!t]
\begin{center}
{\small
\begin{tabular}{@{}lrrllll@{}}
\toprule
                  & \multicolumn{1}{l}{esc50} & \multicolumn{1}{l}{watkins} & \multicolumn{1}{l}{bats} & \multicolumn{1}{l}{cbi} & hbdb                      & dogs                      \\ \midrule
LR (baseline)     & (0.428)                     & (0.776)                       & (0.661)                    & (0.156)                   & (0.751)                    & (0.885)                    \\
IPA               & 0.068                     & 0.040                       & 0.263                    & 0.003                   & 0.273                    & 0.285                    \\
ISPA-A            & 0.206                     & 0.386                       & 0.263                    & 0.004                   & 0.704                    & 0.348                    \\
ISPA-F, MFCC, raw & 0.363                     & {\bf 0.689}                       & 0.469                    & 0.003                   & 0.728                    & {\bf 0.659}                    \\
ISPA-F, MFCC, seg & 0.360                     & 0.652                       & 0.380                    & 0.004                   & 0.734                    & 0.638                    \\
ISPA-F, MFCC, phn & 0.385                     & 0.612                       & 0.387                    & 0.003                   & {\bf 0.741}                    & 0.645                    \\
ISPA-F, AVES, raw & {\bf 0.528}                     & 0.624                       & {\bf 0.491}                    & {\bf 0.159}                   & 0.723                    & 0.580                    \\
ISPA-F, AVES, seg & 0.383                     & 0.454                       & 0.292                    & 0.079                   & 0.723                    & 0.501                    \\
ISPA-F, AVES, phn & 0.383                     & 0.475                       & 0.279                    & 0.081                   & 0.718                    & 0.458                    \\
AVES (topline)    & (0.773)                     & (0.879)                       & (0.748)                    & (0.598)                   & (0.810)                    & (0.950)                   \\ \bottomrule
\end{tabular}
}
\end{center}
\vspace{-1.0em}
\caption{Main results with fine-tuned RoBERTa. Accuracy is an average of three evaluation runs.}
\vspace{-1.0em}
\label{table:results-finetuned}
\end{table*}

In our experiments, we transcribed audio into ISPA-encoded text and assessed its utility in bioacoustic tasks. We used classification datasets from BEANS~\cite{hagiwara2022beans}, a bioacoustics benchmark, in order to evaluate the accuracy of audio classification solely based on its ISPA representations. The following is the list of classification datasets we used:

\begin{itemize}
\setlength\itemsep{-0.5em}
\item {\tt esc50}~\cite{piczak2015dataset} A dataset of environmental audio including animal, nature, and human non-speech sounds.
\item {\tt wtkn}~\cite{sayigh2016watkins} (The Watkins Marine Mammal Sound Database) is a database of marine mammal sounds which contains the recordings of 32 species.
\item {\tt bats}~\cite{prat2017annotated} contains Egyptian fruit bat calls. The target label is the emitter ID (10 individuals).
\item {\tt cbi}~\cite{cornell2020} is from the Cornell Bird Identification competition hoted on Kaggle containing 264 bird species.
\item {\tt dogs}~\cite{yin2004barking} contains dog barks recorded from 10 individual domestic dogs in different situations (disturbance, isolation, and play).
\item {\tt hbdb}~\cite{kiskin2021humbugdb} (HumBugDB) is a collection of wild and cultured mosquito wingbeat sounds. 
\end{itemize}

For ISPA-F, we examined two types of audio feature representations—MFCC (mel-frequency cepstral coefficients) and AVES~\cite{hagiwara2023aves} (the AVES-{\tt bio} model). The feature dimensions for MFCC and AVES are 40 and 768, respectively.

For all the methods above, we feed ISPA-transcribed audio into RoBERTa~\cite{liu2019roberta}, a powerful masked language model (MLM), after tokenization, extract its pooled output, and feed it into a classification head. The entire network is trained using a cross entropy loss with the Adam optimizer, using a learning rate of $1.0\times10^{-4}$ for 15 epochs. The best model based on the validation set was used for testing. 

As mentioned earlier, one of the advantages of ISPA is its versatility, allowing the application of methods developed for NLP as if ISPA were a type of ``foreign language.'' We also present additional results where the RoBERTa model was further fine-tuned using a relatively large dataset of ISPA-transcribed audio in a self-supervised manner. Specifically, we transcribed the entire FSD50K dataset (both the {\tt dev} and {\tt eval} portions) into each ISPA representation and fine-tuned RoBERTa (base) with the MLM objective for three epochs. All other hyperparameters follow the default settings of the {\tt run\_mlm} script from Transformers~\cite{wolf2020transformers}.

\subsection{Main results}

In Table~\ref{table:results-base}, we present the accuracy results for six classification datasets, evaluated using the RoBERTa (base) model without fine-tuning. Notably, transcribing animal sounds as if they were human speech with IPA (indicated in the `IPA' row) demonstrates poor performance, while feature-based methods (ISPA-F) achieve higher accuracy. The table also provides the number of tokens required to represent one second of input audio for each method. Overall, the segmentation and phone-based ISPA-F methods strike a good balance between classification accuracy and conciseness.

Table~\ref{table:results-finetuned} presents the accuracy results obtained by evaluating the fine-tuned RoBERTa model for each method. For comparison, we included both the baseline (logistic regression with 40-dimensional MFCC) and the topline (representing the best results with AVES~\cite{hagiwara2023aves}). The results indicate that fine-tuning the base model proves effective across almost all settings. Remarkably, the performance of certain configurations, particularly ISPA-F with AVES features, approaches or even surpasses that of the baseline method, which uses full, continuous audio features as input. Our experiments show that there's no necessity to rely on audio-specific architectures and pre-training objectives, such as HuBERT~\cite{hsu2021hubert} or AST~\cite{gong2021ast}, to harness the power of large-scale self-supervision. It's worth noting that our fine-tuning experiments are conducted on a relatively small scale (just 50k audio files), and we anticipate observing performance gains as we scale up the model size, data size, and/or compute~\cite{kaplan2020scaling}.

\section{Conclusion}

We introduced ISPA (Inter-Species Phonetic Alphabet), a system designed to transcribe animal sounds into text with the aim of providing an accurate, concise, and interpretable solution. Through our experiments, we demonstrated that certain transcription methods, particularly the feature-based approach with a fine-tuned language model, can achieve performance comparable to ML methods relying on continuous, dense audio representations for bioacoustics classification tasks. While our exploration in this paper focused on classification, the use of text transcription opens up a myriad of possibilities in bioacoustics, including, but not limited to, detection, multimodal processing, audio captioning, and even generation---topics we defer to future work.

\setlength{\bibitemsep}{0\baselineskip}

\bibliographystyle{IEEEbib}
\bibliography{refs_shorter}

\end{document}